\documentclass[12pt]{iopart}

\usepackage{graphicx}
\usepackage{xcolor}
\usepackage{ulem}
\usepackage{verbatim}
\usepackage{hyperref}
\usepackage{upgreek}
\usepackage{soul}
\usepackage{pdfpages}

\newcommand\SoulColor{%
  \let\set@color\beamerorig@set@color
  \let\reset@color\beamerorig@reset@color}
\makeatother

\setstcolor{red}

\begin{document}

\title[Enhancement of the critical current by surface irregularities]{Enhancement of the critical current by surface irregularities in Fe-based superconductors}

\author{I F Llovo$^{1,2,3}$, J Mosqueira$^{1,2}$, Ding Hu$^4$, Huiqian Luo$^5$ and Shiliang Li$^5$}

\address{$^1$ QMatterPhotonics Research Group, Departamento de F\'isica de Part\'iculas, Universidade de Santiago de Compostela, 15782, Santiago de Compostela, Spain}

\address{$^2$ Instituto de Materiais (iMATUS), Universidade de Santiago de Compostela, 15706 Santiago de Compostela, Spain}
\address{$^3$ Centro de Supercomputaci\'on de Galicia (CESGA), 15705, Santiago de Compostela, Spain}

\address{$^4$ School of Physics, Hangzhou Normal University, Hangzhou 311121, PR China}

\address{$^5$ Beijing National Laboratory for Condensed Matter Physics, Institute of Physics, Chinese Academy of Sciences, Beijing 100190, PR China}

\ead{j.mosqueira@usc.es}

\begin{abstract}
The critical current $I_c$ of single crystals of the iron pnictide superconductor BaFe$_2$(As$_{1-x}$P$_x$)$_2$, has been studied through measurements of magnetic hysteresis cycles. We show that the introduction of surface irregularities in the $\upmu$m scale significantly increase $I_c$, primarily near the irreversibility magnetic field $H_{irr}$, where the surface currents are the main contribution to $I_c$. Such an increase is consistent with a theoretical estimate for the maximum non-dissipative current that a rough surface can sustain, based on Mathieu-Simon continuum theory for the vortex state.
\end{abstract}


\submitto{\SUST}

\section{Introduction}
Since the 2008 discovery of superconductivity at high critical temperature in iron-based superconductors (FeSC),\cite{Kamihara08} intensive research on these materials has been taking place. On the one hand, these materials have a fundamental interest, as they share multiple similarities with high-$T_c$ cuprates, such as elevated transition temperatures and the emergence of superconductivity with the introduction of dopants which destroy the antiferromagnetic order of the parent system,\cite{Johnston10,Stewart11,Ishida09} suggesting their pairing mechanism may be similar.\cite{Wang11} 

They also present a  multiband electronic structure,\cite{Johnston10,Stewart11,Ishida09} which leads to unconventional behavior of observables such as the magnetic penetration depth,\cite{Rey14-2} the Seebeck coefficient\cite{Pallecchi16}, the specific heat,\cite{Maksimov11,Hardy10} or the upper critical field.\cite{Hunte08,Gurevich11,Xing17,Llovo21} On the other hand, these materials present some properties that make them interesting from an applications' perspective, such as high critical and irreversibility magnetic fields,\cite{Miura13,Ishida17,Mohan10,Sakoda18,Talantsev19} a relatively low anisotropy. The 122 family of iron-based superconductors is one of the most prominent candidates for next generation superconducting tapes and wires, as the compounds from this family show particularly low anisotropies and large grain boundary critical angles (as high as 9$^\circ$), making them suitable for techniques such as powder-in-tube (PIT) manufacture of wires and tapes,\cite{Katase11,Hosono18,Yao19,Pyon20,Yao21,Zhang22} and for the development of superconducting devices such as bulk magnets\cite{Weiss15} or thin-film nanocircuits including integrated Josephson junctions and SQUIDs.\cite{Hosono18} 

For the aforementioned reasons, much attention has been put on enhancing the critical current density $J_c$ of these materials. Multiple techniques have been shown to serve this purpose, such as proton, neutron and heavy ion irradiation to produce point-like\cite{Eisterer18,Maiorov12,Taen12} or columnar \cite{Eisterer18,Maiorov12,Torsello20,Nakajima10,Otabe12} defects; optimizing the substrate conditions (in the case of thin films),\cite{Sato16,Hiramatsu17} the addition of self-assembled oxygen-rich impurities,\cite{Tarantini12} and the artificial generation of compositionally modulated superlattices, e.g. BaFe$_2$As$_2$/Ba(Fe$_{1-x}$Co$_x$)$_2$As$_2$.\cite{Lee-Tarantini13}

In addition to bulk pinning, surface irregularities have also been shown to be effective at enhancing the critical current in conventional low-$T_c$ superconductors.\cite{Mathieu88,Lazard02} Among the procedures used to create surface defects in these materials are sandblasting,\cite{Lazard02} mechanical abrasion,\cite{Aburas17} buffered chemical or electrolytic polishing,\cite{Casalbuoni05} and low temperature bakeout.\cite{Sung15} More recent work in metallic niobium sheets has shown that laser-induced periodic structuration can make the irreversibility field $H_{irr}$ increase\cite{Cubero20} or decrease when using a femtosecond-pulsed laser.\cite{Cubero20-2} Nevertheless, to our knowledge no similar studies have been conducted on FeSC so far. 

In this work, the effect of surface irregularities on the critical current of FeSC single crystals is studied. The 122 family is the most studied among the FeSC in applied research due to its lower anisotropy and higher $J_c$.\cite{Hosono18} For this reason, we have selected optimally-doped BaFe$_2$(As$_{1-x}$P$_x$)$_2$ (Ba122:P) for this study, as crystals of considerable dimensions and high stoichiometric quality can be grown.\cite{Nakajima12,Hu15,Ramos15} The results will be compared with a theoretical estimate for the critical current density that a rough surface can sustain, based on Mathieu-Simon continuum theory for the mixed state.\cite{Mathieu88,Lazard02}

\begin{figure}[t]
    \centering
    \includegraphics[width=0.5\textwidth]{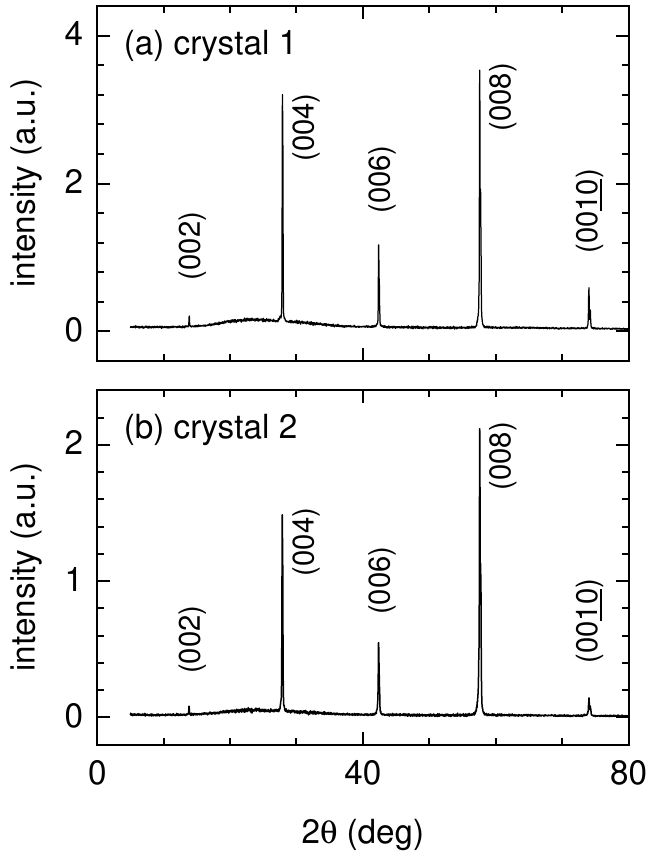}
    \caption{X-ray diffraction patterns of crystal 1 and crystal 2 before and after the treatment, respectively. The data were obtained using the geometry to observe the reflections by the $ab$ layers. Only (00$\ell$) diffraction peaks are present, evidencing the excellent structural quality of the crystals. Moreover, the positions and widths of the peaks are not visibly changed by the surface treatment, indicating that the structural quality of the crystals remains unaltered.}
    \label{FigXRD}
\end{figure}

\begin{figure}[t]
    \centering
    \vspace{1.5pt}
    \includegraphics[width=0.5\textwidth]{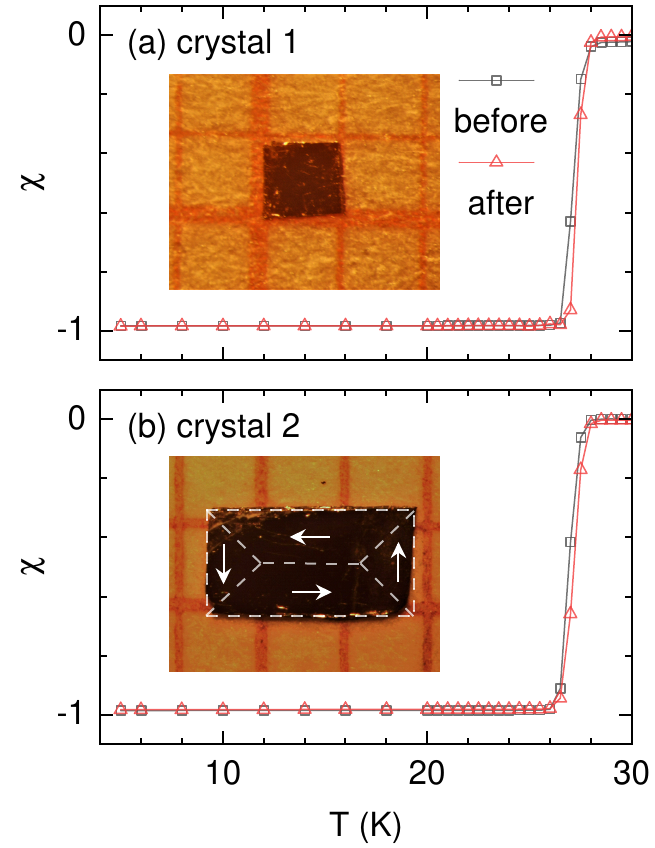}
    \caption{Magnetic susceptibility vs temperature for the BaFe$_2$(As$_{1-x}$P$_x$)$_2$ crystals studied, performed before and after the surface treatment. They have $T_c$ values about 27 K and transition widths under 1 K, evidencing excellent stoichiometric purity, which was not affected by the surface treatment. Photographs of the samples are included in the insets. The arrows on crystal 2 indicate the critical-state electrical current distribution used to evaluate (\ref{Eq:Dmh}).}
    \label{FigTc}
\end{figure}

\section{Methods}
\subsection{Crystals growth and characterization}

BaFe$_2$(As$_{1-x}$P$_x$)$_2$ ($x=0.35$) single crystals were grown using the Ba$_2$As$_3$/Ba$_2$P$_3$ self-flux method described in \cite{Nakajima12}. They are platelike, with typical surfaces of several mm$^2$ and thicknesses up to $\sim$0.1 mm (see Table~\ref{TableSamples}). A thorough characterization of crystals from the same batch was presented in \cite{Ramos15}, where energy dispersive x-ray (EDX) analysis showed an excellent compositional homogeneity, with less stoichiometric deviation than 0.4\% between different crystals and different studied areas. The crystals used in this work are among those used in \cite{Ramos15}, and have been since kept in an epoxy matrix until the beginning of this study. Their crystallographic structure was analyzed by x-ray diffraction (XRD), using a Rigaku MiniFlex II diffractometer with a Cu target and a graphite Cu K$\alpha$ monocromator. The reflections by the crystal $ab$ planes (parallel to the FeAs layers) are presented in figure \ref{FigXRD}. The absence of reflections other than the (00$\ell$) indicates the excellent structural quality of the crystals, and the resulting $c$-axis lattice parameter is about 12.80~\AA~(see table~\ref{TableSamples}), in agreement with data in the literature for crystals with a similar As-P proportion.\cite{Jiang09,Goh10,Ishikado11}.

\begin{figure}[h]
    \centering
    \includegraphics[width=0.5\textwidth]{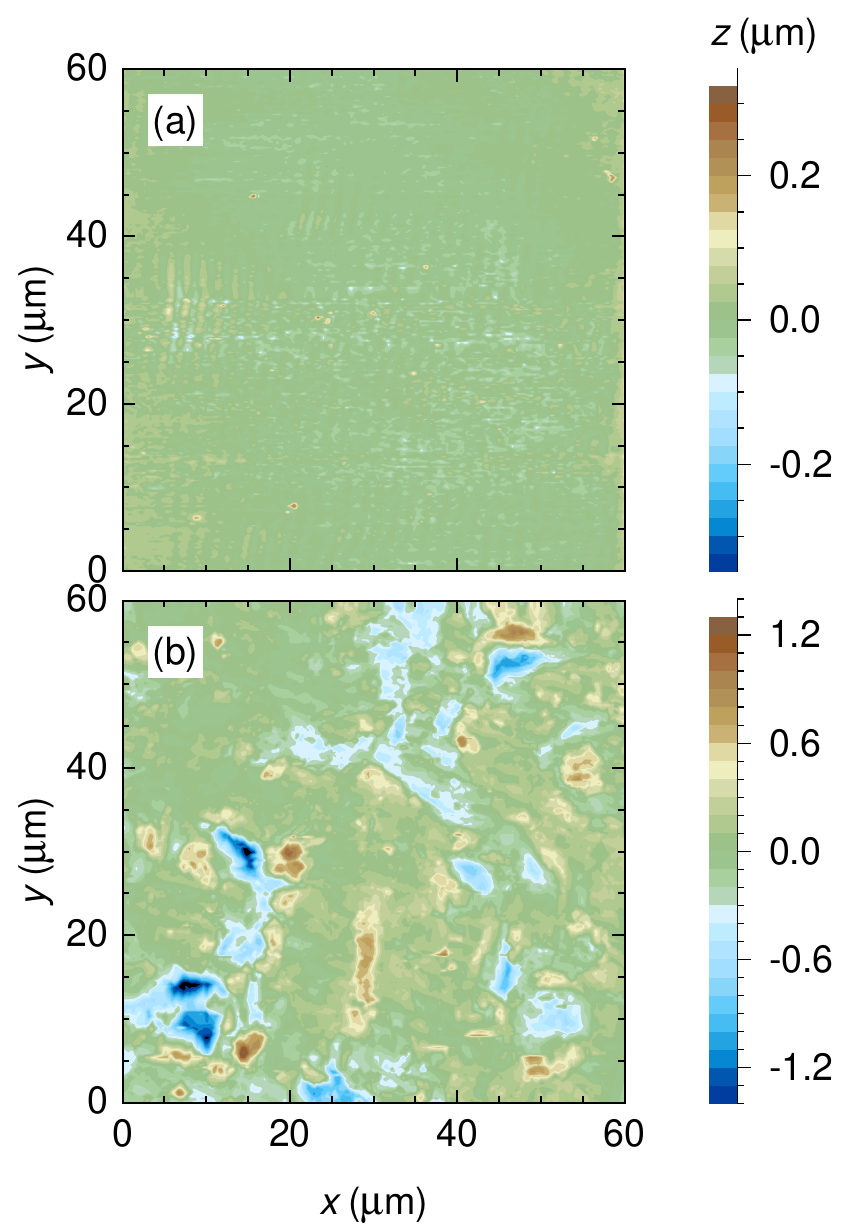}
    \caption{Example of AFM micrographs of the crystal surfaces (a) before and (b) after the abrasive blasting process. While in (a) the irregularities are limited to $\pm0.1\;\mu$m, in (b) they extend up to $\pm1.4\;\mu$m (note the difference in the scale).}
    \label{Fig3AFM}
\end{figure}

\subsection{Surface etching process}

\begin{table}
\caption{\label{TableSamples}Characteristics of the crystals studied. The critical temperature $T_c$ was obtained from low-field magnetic susceptibility measurements (figure~\ref{FigTc}). The $c$-axis lattice parameter $c$ was obtained from x-ray diffraction analysis (figure~\ref{FigXRD}). Both $T_c$ and $c$ agree with data in the literature for BaFe$_2$(As$_{1-x}$P$_x$)$_2$ with the same P~content. The demagnetizing factor for $H\parallel c$,\linebreak $D_c$, was estimated from the dimensions of the samples.}
\begin{indented}
\item[]\begin{tabular}{@{}lllllll}
\br
crystal & $T_c$ & $c$ & size ($L_a\times L_b\times L_c$) & $D_c$ & mass & sides \\ 
  & (K) & (\AA) & (mm$^3$) & & (mg) & etched \\ 
\mr
1 & 27.2  & 12.802(3) & $0.94\times0.94\times0.042$ & 0.933 & 0.230 & 1 \\
2 & 26.9 & 12.795(3) & $2.48\times1.28\times0.070$ & 0.937 & 1.386 & 2 \\ 
\br
\end{tabular}
\end{indented}
\end{table}

The crystals were subjected to an abrasive sandblasting process to create micrometric irregularities on their surfaces. A commercial sandblasting machine (Damglass E. Fexas, DAM-1) was used with silica sand of diameter~$\sim50-100\;\upmu$m. The samples were kept centered in the ejection cone, with the nozzle-to-sample distance set to~8.5~cm. A 5 second burst at 1~bar nozzle pressure was used to etch one of the $ab$ surfaces of crystals 1 and 2. The reverse side of crystal 2 was also subjected to an identical etching process. After the sandblasting, the samples were carefully cleaned from sand residue. Atomic force microscopy (AFM) micrographs of the surface of the samples before and after the sandblasting process are shown in figure~\ref{Fig3AFM}. As it can be seen, the sandblasting process creates defects typically $\sim 5\;\upmu$m wide and $\sim1\;\upmu$m deep, covering most of the surface of the crystals. As shown in figure~\ref{FigXRD}, the XRD peak positions and widths do not change after sandblasting, evidencing that the structural quality of the crystals is preserved after the surface treatment. This is confirmed by extra XRD measurements performed before and after sandblasting on another single crystal (crystal 3) with a higher 2 bar nozzle pressure (see figures~S1 and S2 of the supplementary materials).

\subsection{Magnetization measurements}

The magnetization measurements were performed with a commercial SQUID magnetometer (MPMS-XL, Quantum Design) with the magnetic field applied perpendicular to the $ab$ layers of the crystals. For this purpose, a quartz tube was used as a sample holder. The crystals were placed in a slit perpendicular to the tube axis, and glued with a small amount of GE varnish. Two plastic rods at the tube ends ensured an alignment of the $c$ axis of the crystals with the applied magnetic field of about 0.1$^\circ$.

\begin{figure}
    \centering
    \includegraphics[width=0.7\textwidth]{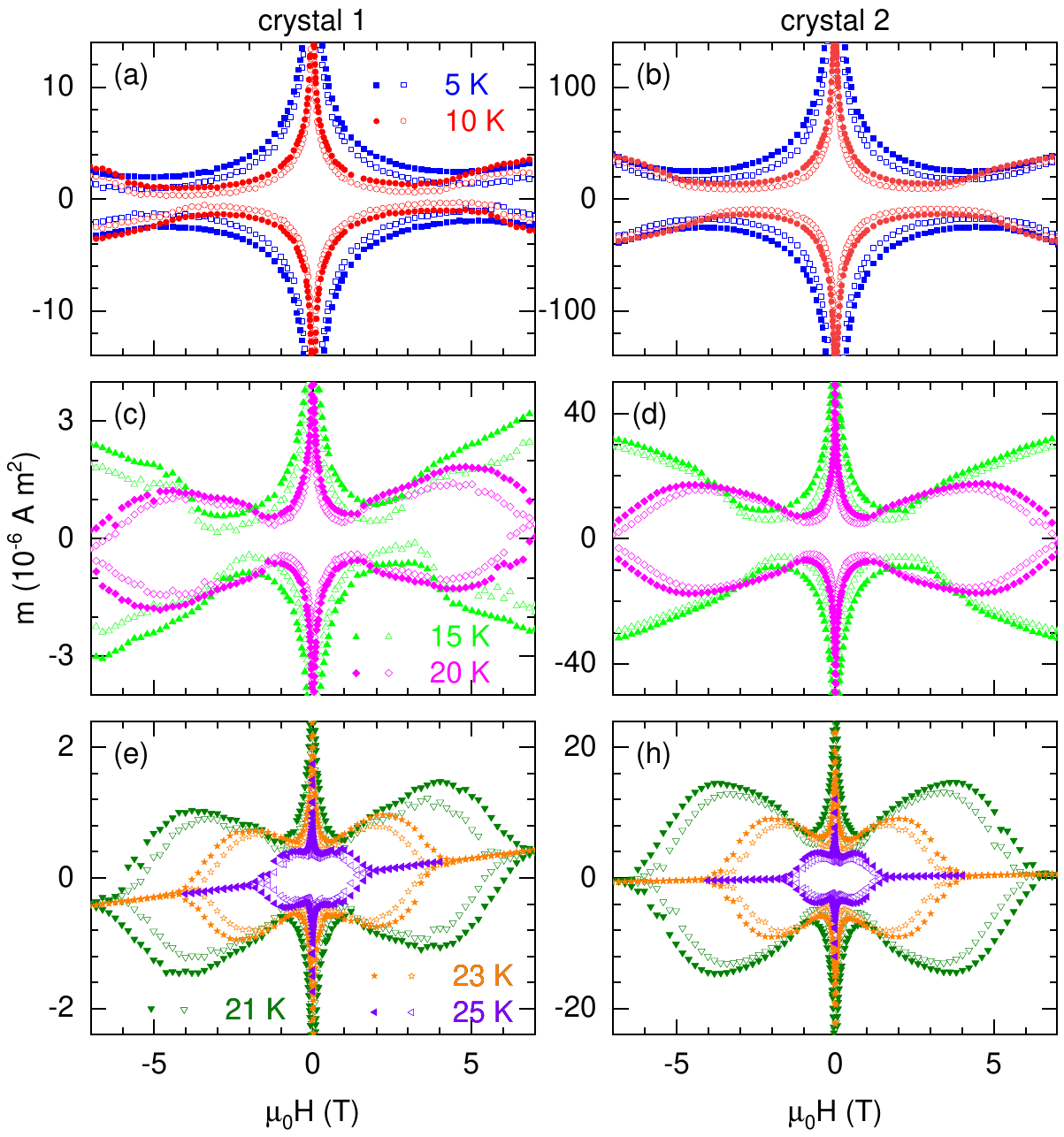}
    \caption{Isothermal $m(H)$ hysteresis curves for the two single crystals, before (open symbols) and after (solid symbols) sandblasting. Crystal 1 was sandblasted only from one side, while crystal 2 was sandblasted from both sides. As it can be seen, the hysteresis amplitude increased after the sandblasting process for both samples.}
    \label{Fig4}
\end{figure}

\section{Results}

The temperature dependence of the magnetic susceptibility $\chi$, measured after zero-field-cooling (ZFC) with a low field ($\sim 0.3$~mT) perpendicular to $ab$ layers, is presented in figure~\ref{FigTc}. These data are corrected for demagnetizing effects by using the demagnetizing factors $D$ calculated from the dimensions of the crystals (see table~\ref{TableSamples}). As shown, $\chi$ is close to the ideal value of -1 at low temperatures, and the diamagnetic transition is very sharp (less than $\sim 1$~K wide), which confirms the excellent stoichiometric quality of the crystals. The superconducting transition temperature, estimated from the diamagnetic transition midpoint, is $T_c\approx 27$~K (see table~\ref{TableSamples}) typical of optimally-doped BaFe$_2$(As$_{1-x}$P$_x$)$_2$.\cite{Chaparro12} It is also worth noting that the surface treatment did not notably affect the diamagnetic transition, confirming that it does not alter the crystals homogeneity.

\begin{figure}[t]
\centering
\includegraphics[width=0.5\textwidth]{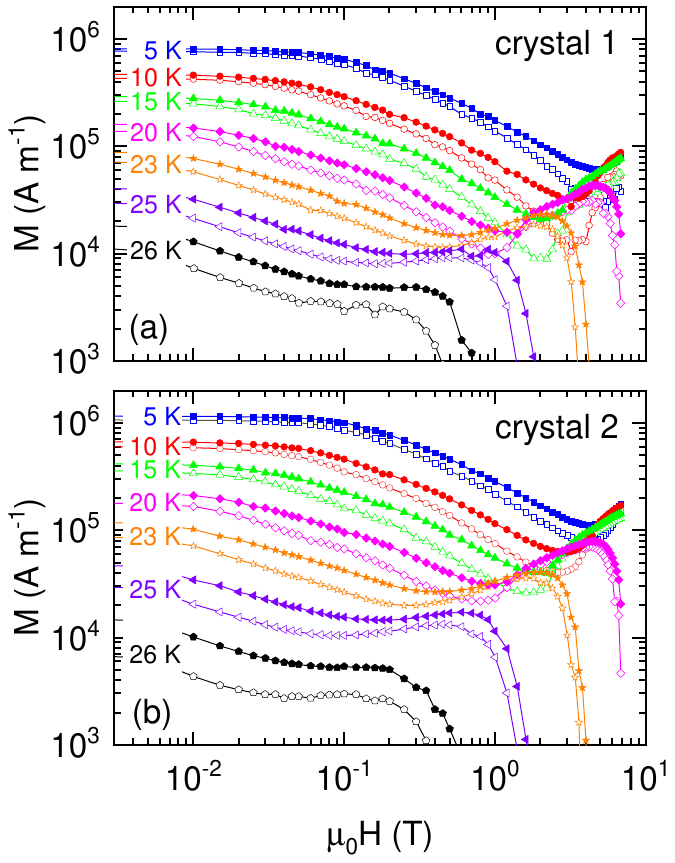}
\caption{Detail of isothermal $M(H)$ cycles for $M,H>0$, before (open symbols) and after sandblasting (solid symbols). The magnetization hysteresis increases at all applied magnetic fields, and is comparatively higher at temperatures closer to $T_c$. The irreversibility field $H_{irr}$ was also increased, suggesting stronger pinning is taking place (see main text for details).}
\label{Fig5}
\end{figure}

To estimate the $ab$-layers' critical current density of the crystals before and after sandblasting, magnetic moment vs. magnetic field $m(H)$ hysteresis cycles were measured with $H\perp ab$ at different temperatures below $T_c$. These measurements were performed by first zero-field cooling (ZFC) to the target temperature. The magnetic field was set using the so-called {\it hysteresis} charging mode (with the power supply continuously turned on), and data were acquired by using MPMS's {\it reciprocating sample option} (RSO), averaging 10~measuring cycles at 1~Hz. Figure~\ref{Fig4} shows the $m(H)$ hysteresis loops obtained for both crystals before (dashed lines) and after (solid lines) sandblasting. As it can be seen, the surface treatment significantly increased the hysteresis amplitude at all the studied temperatures and for all magnetic fields. It is worth noting that this compound presents a {\it second magnetization peak} (SMP), which can be clearly observed for temperatures above 15~K, consistently with previous measurements.\cite{Ishida17,Sugui15,Sugui17}

To observe the effect of sandblasting at low fields and close to $T_c$ more clearly, a detailed log-log representation of the upper-right branch of the hysteresis cycles for both samples is presented in figure~\ref{Fig5}. In order to compare both crystals with each other and with other samples in the literature, $m$ is normalized by the volume of the crystals, and the normal-state paramagnetic signal was subtracted. As it can be seen, the surface treatment produced an increase of the hysteresis amplitude, and this increase extends to the low-$H$ region and is relatively more significant at temperatures closer to $T_c$. Finally, an increase of the irreversibility magnetic field $H_{irr}$ (above which the hysteresis vanishes) was also observed after the sandblasting process. In the following Section, these effects will be interpreted in terms of an extra non-dissipative surface current made possible by the surface irregularities.

\section{Discussion}
\subsection{Critical current enhancement}

\begin{figure}[t]
\centering
\includegraphics[width=0.7\textwidth]{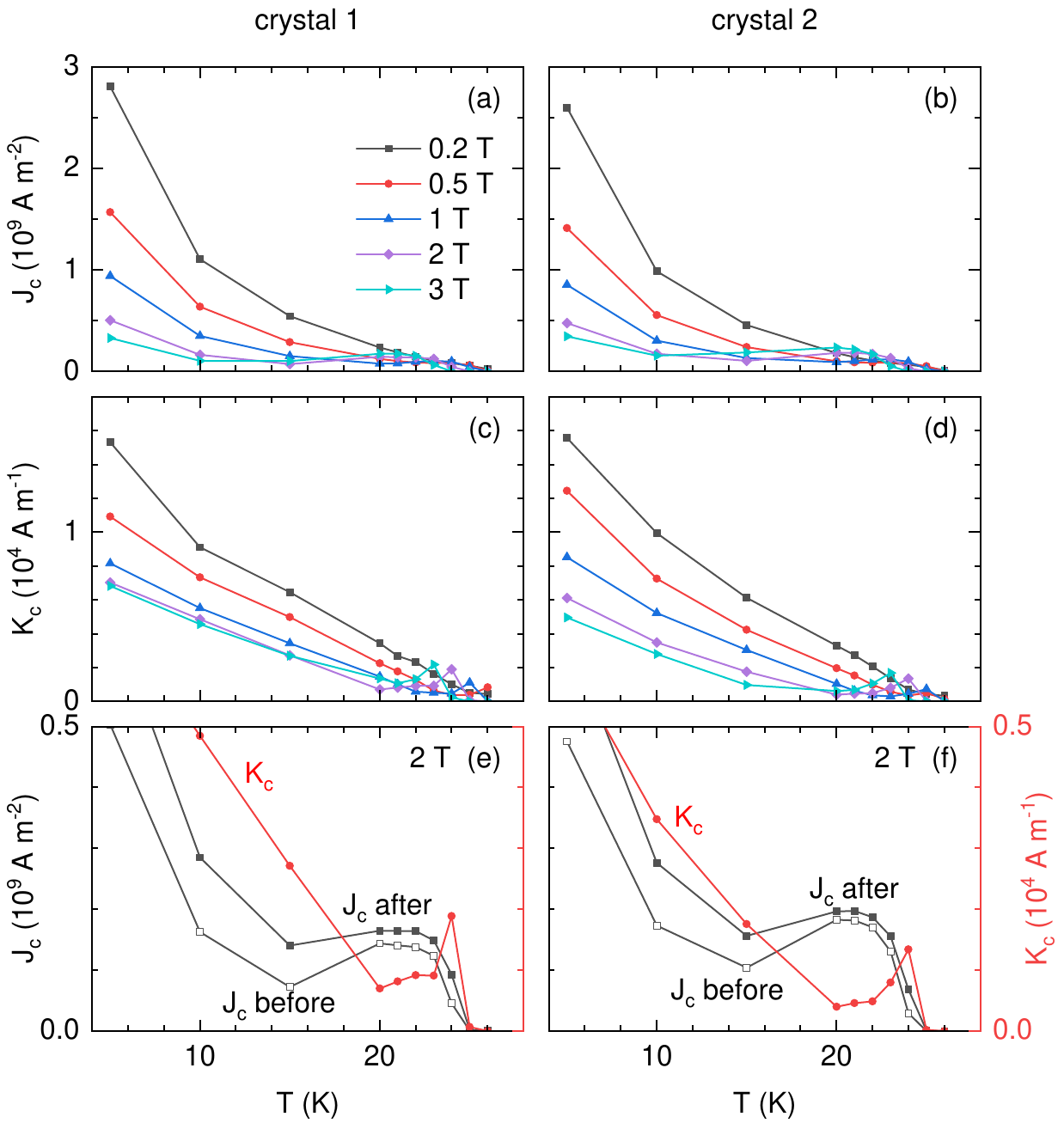}
\caption{(a, b) critical current density $J_c$ before sandblasting, obtained by applying Bean's critical state model; (c, d) critical current surface density $K_c$, obtained by assuming that all additional pinning occurs at the surface of the sample (\ref{Eq:Bean-surface}). As it can be seen, the $K_c(T)$ curves follow a different trend than $J_c(T)$. This can be better observed in (e, f)), where the detail of $J_c(T)$ (black full squares) and $K_c(T)$ (red circles) close to $T_c$ is shown, for an applied magnetic field of 2~T. The empty squares in (e, f) correspond to $J_c(T)$ calculated after sandblasting, with the assumption that no surface currents are present. As shown, the increase is not a constant factor, as it would be expected in first approximation for bulk pinning enhancement. A detail for all measured fields is presented in figure~S3 of the supplementary materials.}
\label{Fig6}
\end{figure}

The $ab$-layers' critical current density $J_c$ before the sandblasting process can be obtained from the $m(H)$ hysteresis loops measured with $H\perp ab$ using Bean's critical-state model\cite{Bean62, Bean64}. For a crystal of dimensions $L_a$, $L_b$ and $L_c$ (where $L_a>L_b>L_c$) it leads to\cite{Poole2nd}
\begin{equation}
J_c = \frac{2m_h/V}{L_b\left(1-L_b/3L_a\right)},
\label{Eq:Jc-Bean}
\end{equation}
where $V=L_aL_bL_c$ and $m_h$ is the amplitude of the $m(H)$ hysteresis loop. The temperature dependence of the resulting $J_c$ at different applied magnetic fields is presented in figures~\ref{Fig6}(a, b) for both crystals. As shown, the $J_c$ curves have a similar amplitude, temperature and magnetic field dependences. 

The increase in the hysteresis amplitude after sandblasting (hereafter denoted as $\Delta m_h$) can be understood as the result of non-dissipative surface currents enabled by surface irregularities. The mechanism for the existence of such currents has been explained in detail in previous works (see, e.g., \cite{Lazard02} and references therein). Let us just mention that the surface roughness gives rise to many ways for the vortices to terminate at the sample surface while satisfying the boundary condition. This enables a collective bending of the vortices which sustains an extra non-dissipative surface current. Defects with size of order $\lambda$ are expected to induce this phenomenon, as $\lambda$ is the characteristic length of the currents around vortices. The magnetic moment $m_s$ due to these surface currents is given by 
\begin{equation}
m_s=\frac{1}{2}\int_{surface}(\vec r\times \vec K)\;d\vec S,
\label{Eq:magnetization}
\end{equation}
where $\vec K$ is the surface current density. In the critical state $\vec K$ corresponds to the critical surface current density $\vec K_c$, and the associated increase in the hysteresis loop amplitude is
\begin{equation}
\Delta m_h=\left|\int_{surface} (\vec r\times \vec K_c)\;d\vec S\right|.
\label{Eq:Dmh}
\end{equation}
To evaluate (\ref{Eq:Dmh}), the spatial dependence of $\vec K_c$ can be approximated by a Bean-like distribution, similar to the one expected for the underlying bulk current density (see diagram inset in figure~\ref{FigTc}(b)). For a crystal of dimensions $L_a$, $L_b$ and $L_c$ (where $L_a>L_b>L_c$) this leads to
\begin{equation}
\Delta m_h = \frac{K_c L_b^3}{2}\left(\frac{L_a}{L_b}-\frac{1}{3}\right)
\label{Eq:Bean-surface}
\end{equation}
per sandblasted surface.

The surface critical current density resulting from (\ref{Eq:Bean-surface}) and the $\Delta m_h$ data in figure~\ref{Fig4}, is shown for both crystals in figures~\ref{Fig6} (c,~d). For crystal 2 the effect of etching both sides was considered. As it can be seen, the result is very similar in both crystals, confirming the existence and reproducibility of the critical current enhancement by surface etching. $K_c$ decreases monotonically with $T$ and $H$, but the $T$ dependence is qualitatively different from the one of $J_c$: while $K_c$ is roughly linear in all the studied temperature range, $J_c$ grows faster at low temperatures. Moreover, $J_c$ presents a broad maximum related to the second magnetization peak (SMP), which is absent in $K_c$ at the same temperature (this can be better seen in the detail near $T_c$ presented in figures~\ref{Fig6}(e,~f). In turn, $K_c(T)$ presents a sharp peak just before vanishing, similar to the {\it peak effect} observed in the critical current of some low-$T_c$ compounds\cite{Bhat94,Xiao00} and high-$T_c$ cuprates,\cite{Kwok94} which has been attributed to the different $T$-dependence of pinning and elastic forces near $H_{c2}(T)$.\cite{Xu08} The different behavior of $J_c$ and $K_c$ confirms that the $m_h$ increase after sandblasting is not due to enhanced bulk pinning but to a true surface effect. 

\subsection{Comparison with theoretical approaches}

\begin{figure}[t]
\centering
\includegraphics[width=0.5\textwidth]{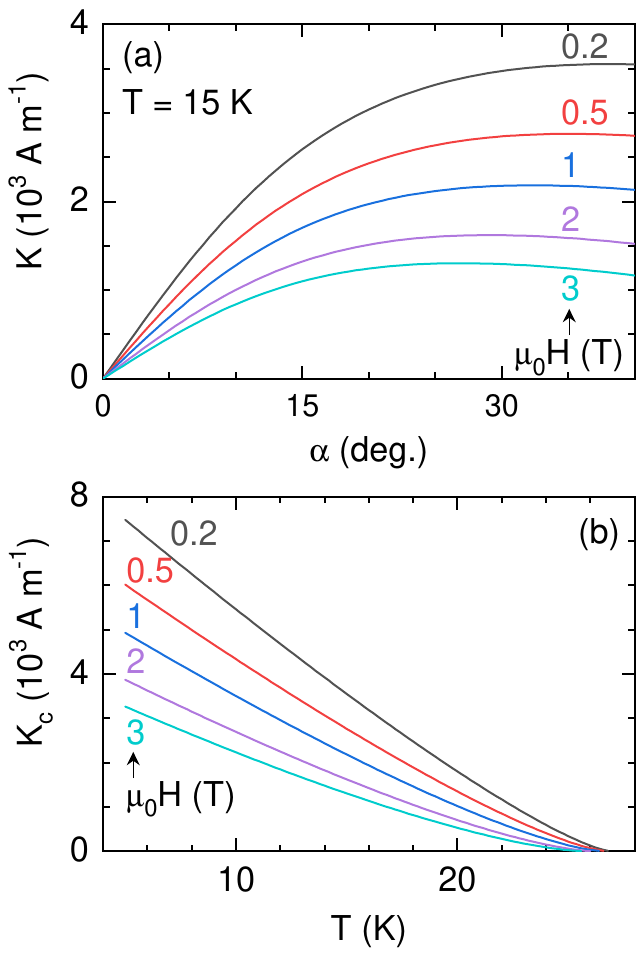}
\caption{(a) Theoretical non-dissipative surface current density against $\alpha$ (the angle between the surface normal and the crystal $c$ axis) for different applied magnetic fields. These curves were obtained from (\ref{Eq:Kalpha}) with the superconducting parameters for optimally-doped BaFe$_2$(As$_{1-x}$P$_x$)$_2$ from \cite{Chaparro12}. As it can be seen, there is an angle $\alpha_0$ for each field for which $K$ is maximum. In a sufficiently rough surface, the critical current density corresponds to $K(\alpha_0)$ (\ref{Eq:Kc-theo}), which is plotted against $T$ in (b). Note the good agreement with the experimental $K_c(T)$ shown in figures~\ref{Fig6} (c, d).}
\label{Fig7}
\end{figure}

We will now discuss if the observed $K_c$ behavior and amplitude is consistent with an estimate based on Mathieu-Simon continuum theory of the mixed state for the non-dissipative current enabled by a rough surface. Let's assume that the external magnetic field $B$ is applied along the $z$ axis (parallel to the crystal $c$ axis), and the electrical current flows in the $y$ direction. At a point in which the surface normal makes an angle $\alpha$ with the $z$ axis, the vortices are bent in the $xz$ plane so that the associated flux density $\omega$ makes an angle $\theta$ with $z$, given by $\tan\theta=\gamma^2\tan\alpha$ ($\gamma$ is the superconducting anisotropy factor). The flux density at the surface is $\omega=B/\cos\theta$. According to Mathieu-Simon theory, the non dissipative local surface current density $K$ is given by\cite{Lazard02}
\begin{equation}
K=|M_x(\omega,\theta)|,
\label{Eq:K-local}
\end{equation}
where $M_x(\omega,\theta)$ is the $x$ component of the reversible magnetization of an anisotropic superconductor under a flux density $\omega$ at an angle $\theta$ relative to the crystal $c$ axis. Using the result from \cite{Kogan88} for the reversible magnetization vector of anisotropic superconductors in intermediate magnetic fields (far from both the upper and lower critical magnetic fields), we find that
\begin{equation}
K=\frac{\phi_0}{8\pi\mu_0\lambda_{ab}^2}\ln\left(\frac{\eta/b}{\sqrt{1+\gamma^2\tan^2\alpha}}\right)\frac{\tan\alpha}{\sqrt{1+\gamma^2\tan^2\alpha}}.
\label{Eq:Kalpha}
\end{equation}
Here, $\lambda_{ab}$ is the magnetic penetration depth for currents along the $ab$ layers, $b\equiv B/B_{c2}^\perp$, where $B_{c2}^\perp$ is the upper critical field perpendicular to the $ab$ layers, $\phi_0$ is the magnetic flux quantum, $\mu_0$ is the vacuum magnetic permeability, and $\eta$ is a constant of order unity. For a given $b$, $K$ has a maximum for some angle $\alpha_0$. This can be seen in figure~\ref{Fig7}(a), where (\ref{Eq:Kalpha}) is plotted against $\alpha$ for different $b$ values, using parameters for optimally-doped BaFe$_2$(As$_{1-x}$P$_x$)$_2$ (see below). If the surface is sufficiently rugous, the vortices have many possible angles to terminate at the surface, and the critical surface current $K_c$ will correspond to (\ref{Eq:Kalpha}) evaluated with $\alpha_0$. The condition $dK/d\alpha|_{\alpha_0}=0$ leads to
\begin{equation}
\gamma^2\tan^2\alpha_0=\frac{1}{2}W\left(\frac{2e^2\eta^2}{b^2}\right)-1,
\label{Eq:alpha0}
\end{equation}
where $W(x)$ is Lambert's function (the inverse of $xe^x$), and $e$ is Euler's constant. Substituting $\alpha_0$ into (\ref{Eq:Kalpha}) and simplifying, we finally obtain
\begin{equation}
K_c=\frac{\phi_0}{16\pi\mu_0\gamma\lambda_{ab}^2}W\left(\frac{2e^2\eta^2}{b^2}\right)\left[1-\frac{2}{W\left(\frac{2e^2\eta^2}{b^2}\right)}\right]^{3/2}.
\label{Eq:Kc-theo}
\end{equation}
A similar calculation was obtained in \cite{Lazard02} using Abrikosov's result for $M_x$ close to the upper critical magnetic field, although this result is not applicable in the present case as the irreversibility field $H_{irr}(T)\sim 0.5H_{c2}(T)$ (see figure~\ref{Fig8}), impeding the study of $K_c$ for fields close to $H_{c2}$. 

Figure~\ref{Fig7}(b) shows the theoretical $K_c(T)$ for the same $B$ values as in figure~\ref{Fig6}, evaluated by assuming a Ginzburg-Landau temperature dependence for the upper critical field $H_{c2}^\perp(T)=H_{c2}^{\perp'}(T-T_c)$ and the magnetic penetration depth $\lambda_{ab}(T)=\lambda_{ab}(0)/\sqrt{1-T/T_c}$. The $\lambda_{ab}(0)$, $H_{c2}^{\perp'}$ and $\gamma$ values used were obtained from from the data in \cite{Chaparro12} for the optimally-doped BaFe$_2$(As$_{1-x}$P$_x$)$_2$. Finally, the parameter $\eta$ was approximated to 1.
As it can be seen, figure~\ref{Fig7}(b) resembles the experimental results for $K_c$ summarized in figures~\ref{Fig6}(c, d), in both the temperature and magnetic field dependences (except for the peak effect observed just before vanishing). The difference in the amplitude (a factor of about two), could be probably attributed to the uncertainties in the superconducting parameters and in the geometry of the samples.

\subsection{$H-T$ phase diagram and irreversibility field increase}

\begin{figure}[h]
\centering
\includegraphics[width=0.5\textwidth]{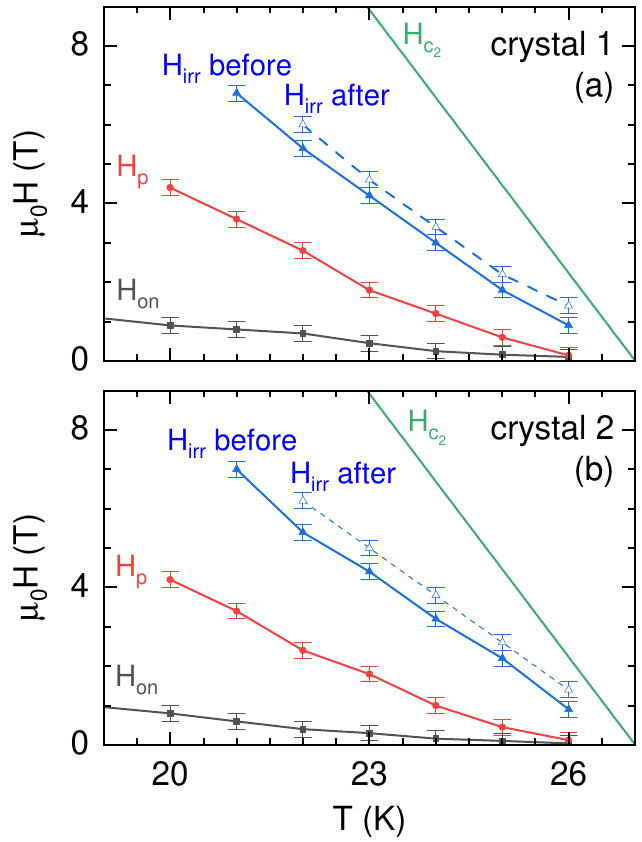}
\caption{$H-T$ vortex phase diagrams for (a) crystal 1 and (b) crystal 2. $H_{on}$ and $H_p$ are, respectively, the onset and the maximum of the SMP before sandblasting. $H_{irr}$ is the irreversibility field before (solid line) and after (dashed line) sandblasting. As a reference, we show the upper critical field line $H_{c2}(T)$, obtained from data in the literature \cite{Chaparro12}.}
\label{Fig8}
\end{figure}

In addition to the critical current enhancement, an increase of the irreversibility field $H_{irr}$ was also observed after sandblasting, as shown in the $H-T$ phase diagrams presented in figure~\ref{Fig8}. The $H_{irr}(T)$ values in this figure were estimated from the $m(H)$ cycles presented in figures~\ref{Fig4}(e,~h), as the magnetic field above which the magnetic moments $m(H^\uparrow)$ and $m(H^\downarrow)$ obtained by increasing and decreasing $H$ respectively (i.e., the upper and lower branch of the hysteresis loops) coincide within the experimental resolution (see figure~S4 of the supplementary materials for an example of this criterion). As it can be seen, the sandblasting process essentially shifts the $H_{irr}(T)$ line to $\sim0.5$~K higher temperatures. Equation (\ref{Eq:Kc-theo}) predicts that $K_c$ should vanish at the $H_{c2}(T)$ line. However, this equation does not consider the effect of thermal fluctuations on the vortices, which play a non negligible role in these materials.\cite{Sugui-Ghivelder09,Mosqueira-Dancausa11,Rey-Carballeira13,Rey14,Ramos15}
For reference, the phase diagrams of figure~\ref{Fig8} also include the onset and the maximum of the second magnetization peak ($H_{on}$ and $H_{p}$, respectively), and the upper critical field line estimated from the data in \cite{Chaparro12}. 

For completeness, figure~S5 of the supplementary materials shows an analysis of the pinning force in terms of Dew-Hughes model using the calculated $H_{irr}$ values, in which no notable differences can be observed after the surface treatment.

\section{Conclusions}

The increase in vortex pinning induced by the addition of surface rugosity through abrasive sandblasting was studied in high-quality single crystals of optimally-doped BaFe$_2$(As$_{1-x}$P$_x$)$_2$. The effect on the critical current was investigated by measuring isothermal $m(H)$ hysteresis cycles with $H$ perpendicular to the etched surfaces (i.e., the $ab$ layers of the crystals), before and after the sandblasting process. A significant increase in the amplitude of the hysteresis loops was observed for both samples and at all temperatures. From this increase, the non-dissipative surface current density $K_c$ that the rough surface can sustain was estimated. $K_c$ presents a temperature dependence qualitatively different from that of $J_c$, estimated from $m(H)$ cycles measured before sandblasting by using Bean's critical state model. However, $K_c$ is in good agreement with both the $T$ and $H$ dependences of a theoretical estimate for intermediate fields based on Mathieu-Simon continuum theory for the mixed state. In addition, a slight increase of the irreversibility field $H_{irr}(T)$ was also observed after sandblasting. Finally, our measurements revealed a sharp increase in $K_c(T)$ just before vanishing (the so called {\it peak effect}), which is not present in $J_c(T)$, suggesting that surface roughness is among the causes that promote its appearance. 

Our findings suggest that surface treatments could be useful to complement the procedures to create bulk pinning, to further enhance the critical current of FeSC wires or tapes for electrical transport applications fabricated using the PIT method.\cite{Hosono18,Yao21,Zhang22} The feasibility of adding surface irregularities could be tested e.g., by chemical or mechanical etching of the inner walls of the tube before introducing the superconducting material. 

\section*{Data availability statement}
The data that support the findings of this study are available upon reasonable request from the authors.

\ack
This work was supported by the Agencia Estatal de Investigaci\'on (AEI) through project PID2019-104296GB-I00. I.F. Llovo acknowledges financial support from Xunta de Galicia through grant ED481A-2020/149. H. Q. Luo is supported by the National Key Research and Development Program of China (Grant No. 2018YFA0704200), the Strategic Priority Research Program (B) of the CAS (Grants Nos. XDB25000000), and the Youth Innovation Promotion Association of CAS (Grant No. Y202001). The research at HangZhou Normal University is supported by the Open Project of Guangdong Provincial Key Laboratory of Magnetoelectric Physics and Devices (Grant No. 2022B1212010008), Startup Project of HangZhou Normal University (Grant No. 2020QDL026) and Natural Science Foundation of Zhejiang Province (Grant No. LY22A040009). Authors would like to thank the use of RIAIDT-USC analytical facilities.

\section*{References}

\bibliography{bibliography}



\section*{Supplementary material}

# Supplementary Information

Enhancement of the critical current by surface irregularities in Fe-based superconductors

I. F. Llovo[a,b,c], J. Mosqueira[a,b,*], Ding Hu[d], Huiqian Luo[e], Shiliang Li[e]

[a]QMatterPhotonics Research Group, Departamento de Física de Partículas, Universidade de Santiago de Compostela, 15782, Santiago de Compostela, Spain
[b]Instituto de Materiais (iMATUS), Universidade de Santiago de Compostela, 15706 Santiago de Compostela, Spain
[c]Centro de Supercomputación de Galicia (CESGA), 15705, Santiago de Compostela, Spain
[d]School of Physics, Hangzhou Normal University, Hangzhou 311121, PR China
[e]Beijing National Laboratory for Condensed Matter Physics, Institute of Physics, Chinese Academy of Sciences, Beijing 100190, PR China

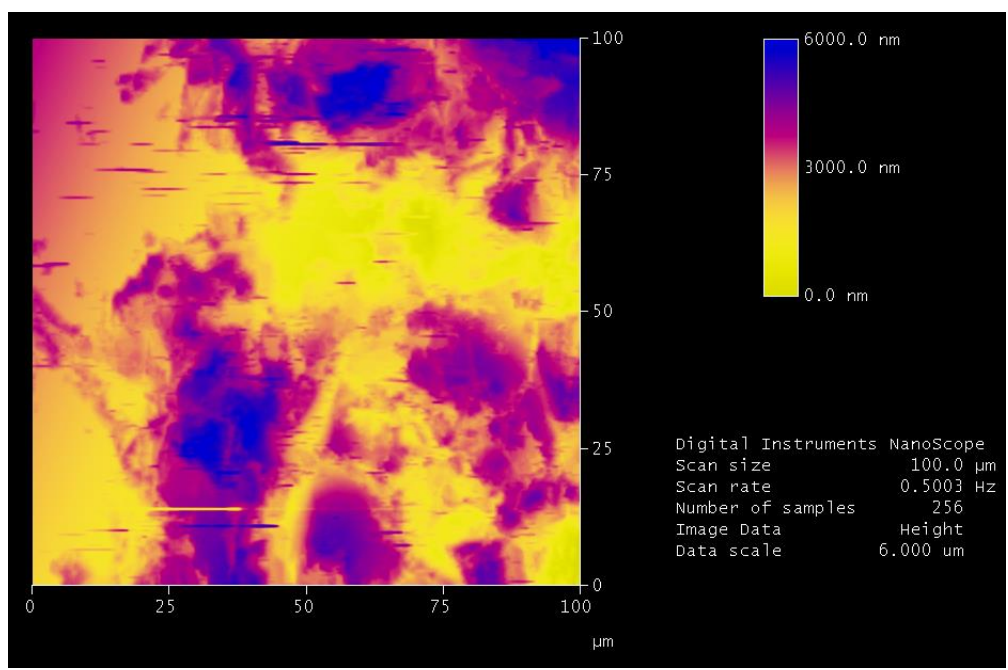

Fig. S1. Atomic-force microscopy (AFM) image of crystal 3 (from the same batch as crystals 2 and 3 used in the magnetization experiments) after sandblasting by using the procedure described in the main paper. The larger nozzle pressure used (2 bar) leads to surface irregularities ± 3 $\mu$m in depth, typically two times larger than the ones in crystals 1 and 2 (sandblasted with 1 bar nozzle pressure). Despite that, the x-ray diffraction measurements presented in Fig. S2 confirm that the surface treatment does not appreciably affect the structural quality of the crystals.

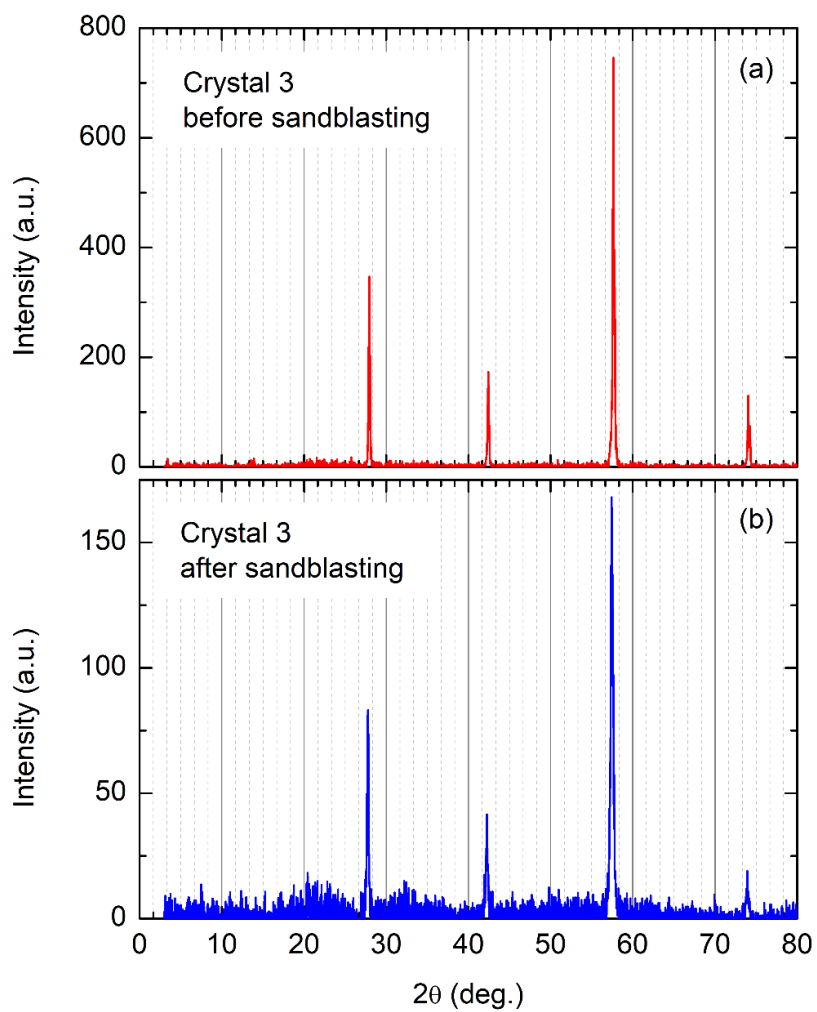

Fig. S2. X-ray diffraction patterns of crystal 3 before (a) and after (b) the sandblasting process in one of its faces. They have been obtained by using the geometry to observe the reflections by the crystal ab planes. The face exposed to the x-rays after sandblasting was the etched one. As shown, the diffraction peak positions and widths are unchanged after the surface treatment, confirming that the structural quality of the crystals has been preserved.

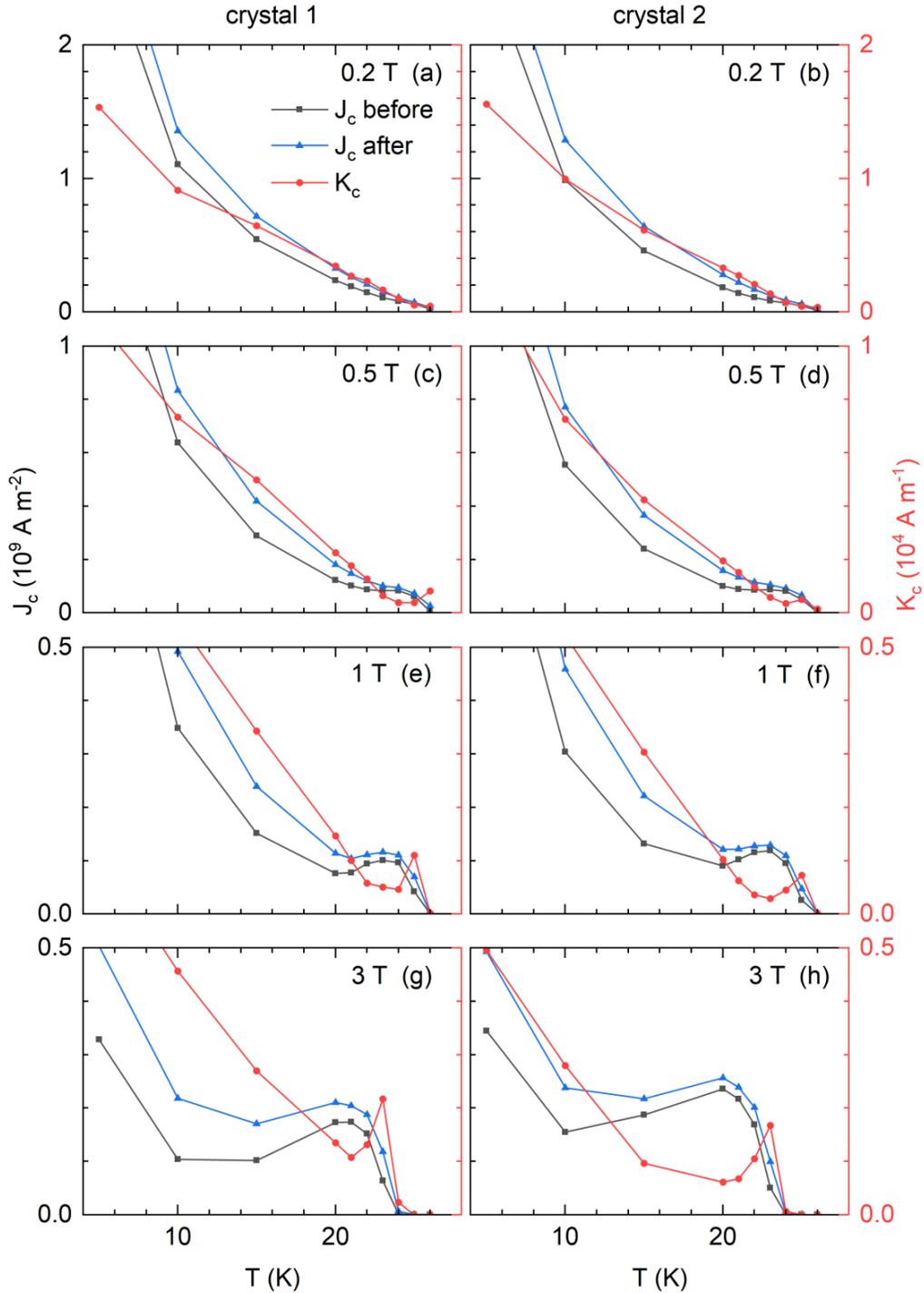

Fig. S3: Comparison of $J_c$ obtained from the m(H) curves before sandblasting (squares, black lines), and the $J_c$ that would result after sandblasting (triangles, blue lines), assuming that a surface critical current is not present. As it can be seen, the $J_c$ increase is not a constant factor, as it would be expected in first approximation for bulk pinning enhancement. The circles/red lines correspond to the surface critical current, $K_c$, calculated from the m(H) hysteresis increase as indicated in the main text. $K_c$ follows a trend significantly different from the one of $J_c$ (see the main text for details).

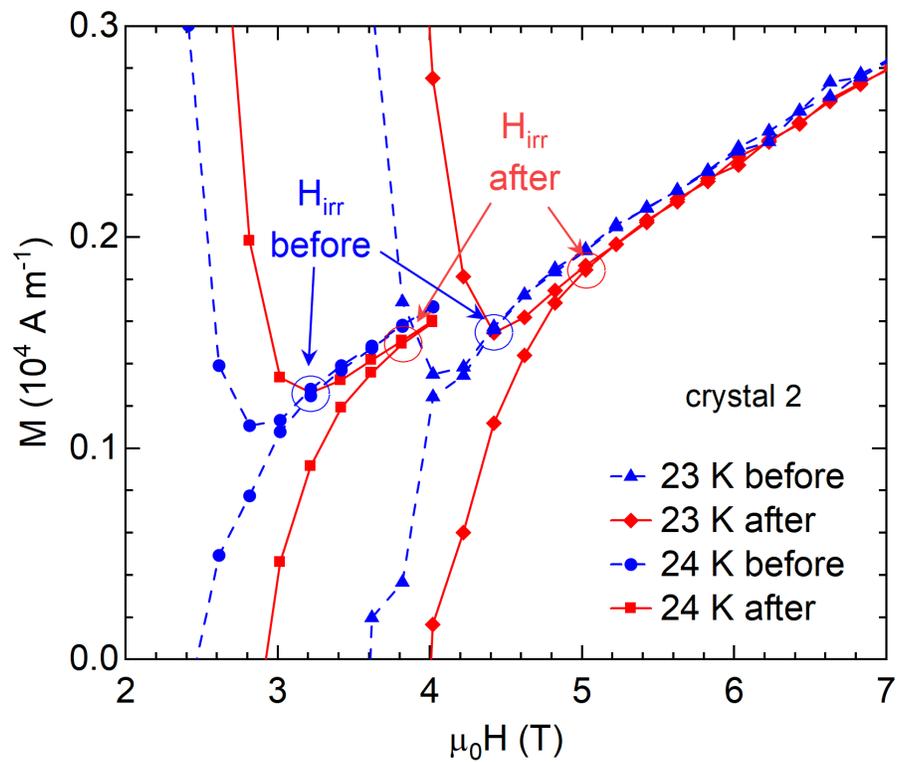

Fig. S4: Criterion used for the calculation of the irreversibility field (see main text for details). This figure clearly shows the $H_{irr}$ increase after sandblasting.

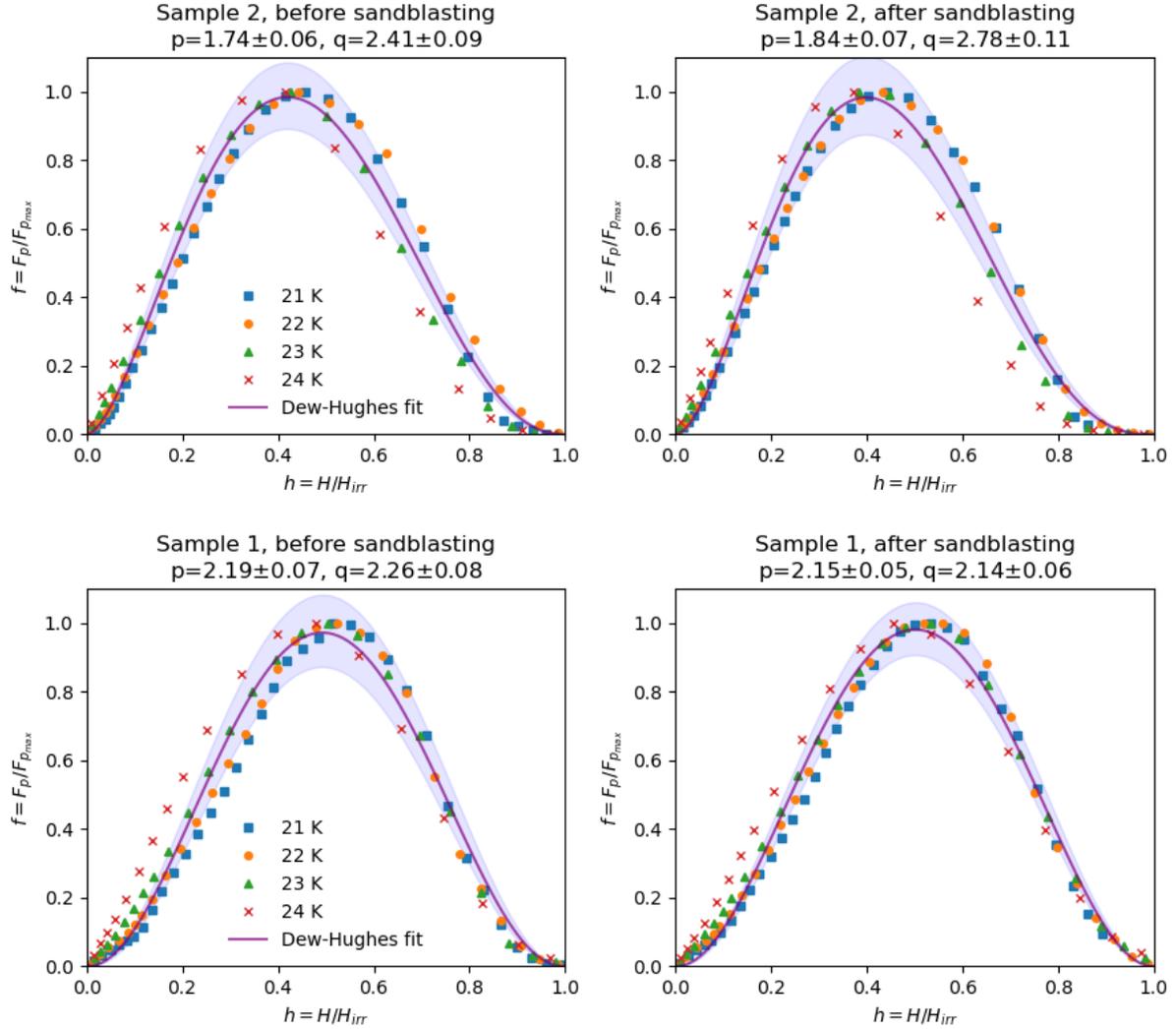

Fig. S5: Normalized pinning force $f = F_p/F_p^{max}$ vs. reduced applied magnetic field $h = H/H_{irr}$, where the pinning force density is obtained as $F_p = J_c B_{app}$. A scaling of the curves for different temperatures is observed, and the differences between the data before and after sandblasting are within the uncertainty of the scaling (indicated as a shadow area). The lines are the best fits to Dew-Hughes model, $f = A\,h^p(1-h)^q$. The value of $h_{max} = p/(p+q)$ is $0.492 \pm 0.018$ before and $0.501 \pm 0.014$ after sandblasting for crystal 1, and $0.420\pm0.016$ before and $0.398\pm0.015$ after sandblasting for crystal 2, consistent with previous very detailed measurements for BaFe$_2$As$_{2-x}$P$_x$ with x=0.30-0.33 (see Fig. 11 in Ishida et al., Phys. Rev. B **95**, 014517 (2017)). The parameters $p$ and $q$, indicated in the figures, are also close to the ones found in BaFe$_{2-x}$Ni$_x$As$_2$ near optimal-doping:

- K. S. Pervakov et al., Supercond. Sci. Technol. **26**, 015008 (2013).
  BaFe$_{1.9}$Ni$_{0.1}$As$_2$, p=1.53, q=2.25, h$_{max}$=0.4

- S. Sundar et al., ACS Appl. Electron. Mater. **1**, 179 (2019).
  BaFe$_{2-x}$Ni$_x$As$_2$ (x=0.108), p=1.5, q=3.8, h$_{max}$=0.28

- V. A. Vlasenko et al., Physics Procedia **67**, 952 (2015).
  BaFe$_{2-x}$Ni$_x$As$_2$ (x=0.09), p=1.6, q=1.8, h$_{max}$=0.48



\end{document}